# Enhancing security of optical cryptosystem with position-multiplexing and ultra-broadband illumination

Dongliang Tang,[1,†] Sujit Kumar Sahoo,[1,2,†,*] Cuong Dang[1,*]

[1] Centre for OptoElectronics and Biophotonics (OPTIMUS), School of Electrical and Electronic Engineering, The Photonic Institute (TPI), Nanyang Technological University Singapore, 50 Nanyang Avenue, 639798, Singapore
[2] Department of Statistics and Applied Probability, National University of Singapore, Singapore
*Corresponding author: sujit@pmail.ntu.edu.sg, hcdang@ntu.edu.sg
†Co-first authors with equal contribution

**A position-multiplexing based cryptosystem is proposed to enhance the information security with an ultra-broadband illumination. The simplified optical encryption system only contains one diffuser acting as the random phase mask (RPM). Light coming from a plaintext passes through this RPM and generates the corresponding ciphertext on a camera. The proposed system effectively reduces problems of misalignment and coherent noise that are found in the coherent illumination. Here, the use of ultra-broadband illumination has the advantage of making a strong scattering and complex ciphertext by reducing the speckle contrast. Reduction of the ciphertext size further increases the strength of the ciphering. The unique spatial keys are utilized for the individual decryption as the plaintext locates at different spatial positions, and a complete decrypted image could be concatenated with high fidelity. Benefiting from the ultra-broadband illumination and position-multiplexing, the information of interest is scrambled together in a small ciphertext. Only the authorized user can decrypt this information with the correct keys. Therefore, a high performance security for a cryptosystem could be achieved.**

http://dx.doi.org/10.1364/OL.99.09999

As the development of the computing science and information technology, the information safety has become more challenging and drawn a lot of attention in recent years. Optical encryption technology has been commonly investigated due to the advent of parallel signal processing, multi-dimensional operations and increasing computational power [1-12]. The pioneer work of the double random phase encoding (DRPE) for optical encryption technology was proposed by Refregier and Javidi in 1995 [1]. Since then many extended optical encryption algorithms and schemes, such as fractional Fourier domain [3, 4] and Fresnel domain [5], have been reported to improve the security strength and enlarge the storage capacity. Similar to the original DRPE algorithm, these methods utilize two independent RPMs as the security key to convert the plaintext into a stationary and seemingly white noise. However, the use coherent illumination in most of these methods is a drawback. Not only encryption system, any optical systems based on coherent illumination are highly sensitive to the optical misalignment and unavoidable coherent artifact noise. To avoid these problems, some interesting technologies which originally were established with coherent illumination, have been redeveloped for the incoherent illumination, such as Fresnel incoherent correlation holography [13, 14], incoherent digital holographic adaptive optics [15], some incoherent optical correlators [16, 17].

Similarly, incoherent illumination has also been utilized for optical cryptosystems [18-21]. Use of a simple optical diffuser as RPM was proposed to greatly reduce the complexity of the system and to decrease the errors generated from the coherent artifact noise [18, 19]. Like DRPE and other optical cryptosystems, the incoherent illumination based optical cryptosystems are prone to ciphertext-only attack (COA) [22], known-plaintext attack (KPA) [23, 24], chosen-plaintext attack (CPA) [25, 26], brute force attack [26], and chosen-cyphertext attack (CCA) [27], etc. Among these, COA is the hardest attack, which requires decryption of the ciphertext without any additional information. However, COA becomes easier in diffuser based optical cryptosystems, because the ciphertext's autocorrelation is essentially similar to the plaintext's autocorrelation. One could recover the plaintext without knowledge of the security key by employing the phase-retrieval algorithm [28]. The basic principle relies on the optical memory effect, which states that light from nearby points on the object will generate nearly identical but shifted random speckle patterns on the other side of a scattering medium. The autocorrelation of these speckle patterns is an impulse function [29]. Hence, the autocorrelation of the object within the memory effect region is preserved through the scattering medium. Increasing security of the incoherent illumination based optical cryptosystems is crucial.

In this letter, we propose a position-multiplexing technique to improve the information security with an ultra-broadband illumination. Our simple optical image encryption setup contains only one RPM to scatter light coming from various spatial objects (i.e. plaintexts) and generate a scrambled speckle pattern (i.e. ciphertext) on the camera when using an ultra-broadband illumination. This ultra-broadband spectrum would totally affect the performance of the previous COA technique because the illumination bandwidth is significantly larger than the RPM's speckle correlation spectral bandwidth. The speckle patterns produced by multiple wavelengths in this very large bandwidth could not stay correlated [29-31]. Therefore, it would break the condition of nearly equivalent autocorrelation between the plaintext and its ciphertext. Hence, it reduces the security risk from COA in the previous incoherent cryptosystems.

To further strengthen the ciphering, we have used the concept of the position multiplexing [32, 33]. The image produced by the RPM is the convolution of the object with the point's speckle pattern or incoherent point spreading function (PSF). The method will work when the PSF is shift-invariant, i.e. the object is within the memory effect region of the scattering medium [29]. Each pixel of the output image contains the object information multiplexed in a random way. Therefore, we make ciphertext smaller than the full object by taking only the center portion of the output image. Such small size of the speckle pattern will make it impossible to estimate the plaintext's autocorrelation for COA. As the information of spatial plaintexts is mixed and speckles from multiple wavelengths are overlapped, only the authorized user with correct spatial keys could decrypt corresponding pieces of information. Therefore, one could obtain a higher security for the text ciphering.

The basic principle of the position-multiplexing cryptosystem is descripted below and presented in Fig. 1. The complete encryption setup is schematically shown in Fig. 1(a). The desired plaintext is displayed on the projector and projected at the input plane, where the iris 1 is used for selecting the plaintext and minimizing the background light from the projector. One diffuser (Edmund, 120 Grit Ground Glass Diffuser) is placed at a distance from the input plane and scrambles the original light field of the plaintext. A scientific camera (Andor Neo 5.5, 2560*2160, pixel size 6.5um) is used to capture the encrypted image at the output plane. Iris 2 with about 2 mm diameter is used for obtaining an appropriate speckle intensity, grain size and signal-to-noise contrast according to the reported reference about the scattering media [29]. The distance from the iris 1 to RPM and distance from RPM to the camera are $u\simeq210mm$ and $v\simeq87.5mm$, respectively. We use full white-light, from 400 nm to 720 nm wavelength, or full green spectrum, from 460 nm to 610 nm wavelength, of the projector in our experiments. Under this ultra-broadband incoherent illumination, the encryption process from the input to the output could be mathematically expressed as the following formula:

$$I(x,y) = O(x,y) * PSF(x,y), \quad (1)$$

where the intensity image $I$ or ciphertext is a speckle pattern as the output, $O$ is a demagnified object intensity or plaintext as the input, the PSF plays the role of security key, $*$ is the convolution operation, and $x$ or $y$ corresponds to the coordinate along $x$ or $y$ direction at the output plane. Equation (1) is similar to the operation in the reported incoherent encryption in Ref. (19), which could be approximated with discrete convolution in the pixel space as

$$I(x,y) = \sum_{i,j} O(i,j) PSF(x-i, y-j). \quad (2)$$

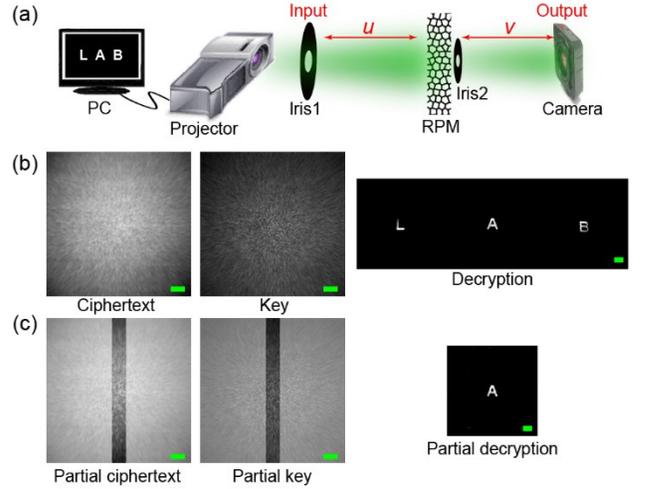

**Fig. 1.** Experimental demonstration of the position-multiplexing principle in a cryptosystem with one RPM. (a) A plaintext with 'LAB' letters at different spatial positions are encrypted through one RPM and a camera records the mixed/scattered ciphertext. (b) Ciphertext, key and decryption as utilizing full camera image for decryption processing. (c) Partial ciphertext, key and decryption as using the reduced images. Scale bars: 200 pixels in ciphertexts and keys, and 20 pixels in decryptions.

The demagnified plaintext could be reconstructed with the correct security key through the deconvolution process as

$$O = FT^{-1}[FT(I)/FT(PSF)], \quad (3)$$

where $FT$ and $FT^{-1}$ are the Fourier transform operation and the inverse Fourier transform operation respectively. In this work, we utilize the standard Wiener deconvolution algorithm to do decryption processing. A reconstruction with higher fidelity could be obtained by using a larger images size, as the reconstruction artefact increases with higher correlation between the noise and actual signal. This noise – signal correlation decreases as the total pixels or the image dimension increases. Therefore, there exist a trade-off between the image quality and the ciphering strength. For clarity, we remove the background which is set at 20% of the maximum intensity. In Fig. 1(b), a decrypted result is shown by taking the intensity $I$ with 2048x2048 pixels as the ciphertext, and the PSF with the same size as the key. In Fig. 1(c), the decrypted result is shown by taking the intensity $I$ with only 200x2048 central pixels as the ciphertext and the same pixel area on PSF as the key. It shows that by cropping the intensity image and PSF, the peripheral objects are completely lost. Even though the partial intensity image $I$ contains every information of the input object, the partial PSF could only be able to reconstruct the object specific to its position and dimension.

The deconvolution is successful because the memory effect of scattering media makes the shift-invariant PSF. In addition, the spatial correlation of different points is similar to that of their corresponding PSFs, because of another interesting property that the speckles (size, shape, intensity and position) on a single PSF are independent from each other. In other words, non-overlapped areas on a single PSF are uncorrelated to each other. As a result, the reconstruction with partial PSF in Fig. 1(c) has no impression of the peripheral text. In order to demonstrate this spatial decorrelation property of the PSF, we have plotted the correlation curve taking various window sizes. We first extract a window (dash-lined rectangle) from the center of the PSF and use it as the reference to compute the correlation with the windows at various shift positions as shown in Fig. 2(a). The relationship between correlation coefficient with pixel shift is plotted in Fig. 2(b). It shows a complete decorrelation of the PSF window when there is no spatial overlaps (i.e. the shift is beyond the window size). Figure 1(c) has also demonstrated this decorrelation, where the letter "A" is clearly reconstructed without any cross-talk (or interference) from "L" and "B". It is because the letters are more pixels apart from the width of the key, which will avoid the cross-talk between the keys.

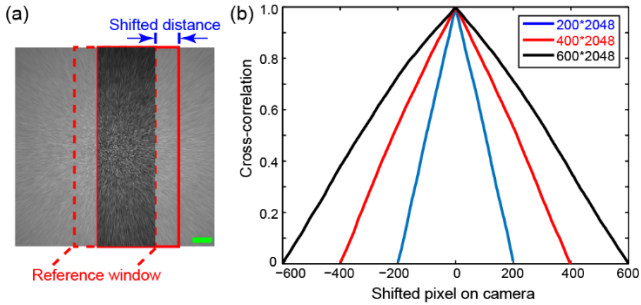

**Fig. 2.** The cross-correlation between the reference window and the windows extracted at different shifts. (a) The measured speckle pattern of the PSF. The transparent region illustrates the correlated speckles between reference window and window extracted at a shifted position (the solid lined rectangle). (b) Cross-correlation coefficient between the extracted windows at different shifts for various window sizes: 200*2048, 400*2048 and 600*2048 pixels. Scale bar: 200 pixels.

The spatial decorrelation of the PSF enables the position-multiplexing of the objects spaced apart as the dimension of the keys. Idea of position-multiplexing can easily be perceived from the discrete convolution formula in Eq. (2), where speckle image is the shifted and weighted sum of the PSF. The spatial decorrelation of the PSF allows us to demultiplex the objects at various spatial positions from the same partial intensity image $I$, which is the advantage in our cryptosystem. Thus, we use the spatially non-overlapping windows of the PSF as the security keys to extract the information of the objects at the corresponding spatial positions. Our position-multiplexing technique can be considered as a method to superpose the cyphertexts of different single RPM based encryption system. We demonstrate this position-demultiplexing capability by extracting 3 non-overlapping keys with size 200x2048 pixels from the PSF, which is shown in Fig. 3(b). Here, the ciphertext is still the intensity $I$ with central 200x2048 pixels, as shown in Fig. 3(a), which has the text information of all the spatial positions in the input plane. Three images are reconstructed from the ciphertext using 3 different keys, and placed at the respective spatial position of the keys. Then this concatenated image is shown in Fig. 3(c) with a perfect recovery of the 3 letters. In practice, the sender and the receiver have already exchanged the keys, and by using these keys, many ciphertext can be transferred and decrypted. The receiver doesn't need to own the RPM or any optical setup. The decryption process is done digitally.

We select the centre portion of the intensity image as a ciphertext,

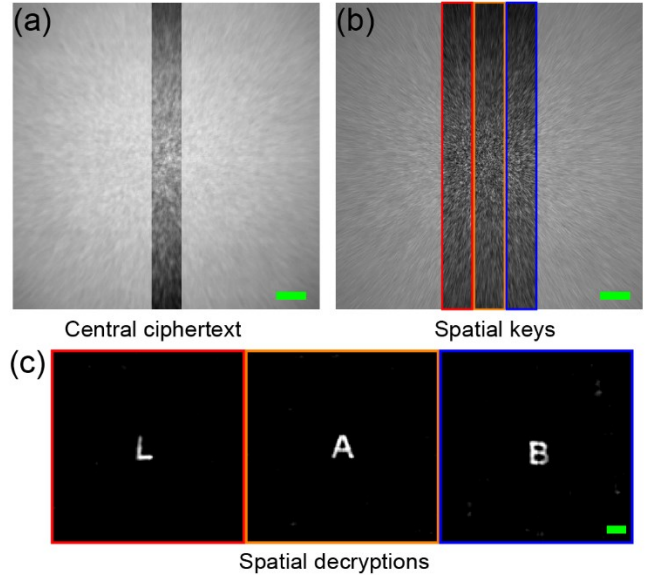

**Fig. 3.** Stitched 1D decryption with various spatial keys. (a) Ciphertext with 200*2048 pixels in central part. (b) Various spatial keys with the same size in (a). (c) A full 1D recovery is stitched together with different spatial decryptions. Scale bars: 200 pixels in (a) and (b), and 20 pixels in (c).

because it has the highest signal-to-noise ratio due to the hallow effect of the speckle pattern. However, in principle, all the pixels in intensity image carry all the object information as a result of convolution. We don't necessarily need to have the keys aligned with the centre intensity image. In the next demonstration, we take a text object having four numbers at four corners. The central area (220*220 pixels) of the intensity image is taken as the ciphertext, which is shown in Fig. 4(a). Then we have taken the same central 220*220 pixels of the PSF as the key, which is shown in Fig. 4(b). The decryption is shown in Fig. 4(c), which shows that central key carries no information as the central portion of the object is blank. Next, we generated four keys as spatial PSFs by dividing the central 440*440 pixels of the PSF, which are displayed in Fig. 4(d). The stitched decryption is shown in Fig. 4(e), which shows a perfect reconstruction of the four letters at the four corners, each with the corresponding key. In this demonstration, the background with the normalized intensity smaller than 0.4 are removed as the reduced sizes for the ciphertext and key are much smaller than in Fig. 1 and generate more reconstruction noises. To confirm the strength of our ciphering method, we run phase retrieval algorithm, which implements hybrid input-output, HIO, and error reduction strategy, with the inputs of ciphertext in Fig. 1b, 3a and 4a. None of them produces any information about the plaintexts.

Both single and double RPM based optical crypto systems falls into the category of linear systems. Linear crypto system are prone to various forms of attacks including COA. Certainly, nonlinearity of the system is critical for information security. However, emphasis of this work is on enhancing the security of linear optical cryptosystem. The COA is mainly attributed to the high contrast of the ciphertext (or the speckle images), as it relies on the accurate estimation of the plaintext's autocorrelation. Our proposed method would increase the security by reducing the speckle contrast with ultra-broadband illumination. In addition, we have significantly reduced the ciphertext size by using the concept of position-multiplexing to make the attack more challenging. Other attacks require deriving the encryption key, such as known-plaintext attack (KPA), chosen-plaintext attack (CPA), chosen-cyphertext attack (CCA) and brute force attack. In contrast with other linear optical encryption methods, our position-multiplexing technique allows us to have a single cyphertext, which carries multiple plaintexts with different keys. This will make it significantly difficult for attacker to derive all the keys with the knowledge of some plaintexts and cyphertexts. The information of multiplexing positions for these keys will add another level of security to our system. The more number of position-multiplexing will enhance the security of the system. At the same time, it will also reduce the number of pixels for the keys and cyphertext, resulting in poor recovery via deconvolution. This sets the tradeoff between the reconstruction quality and the security.

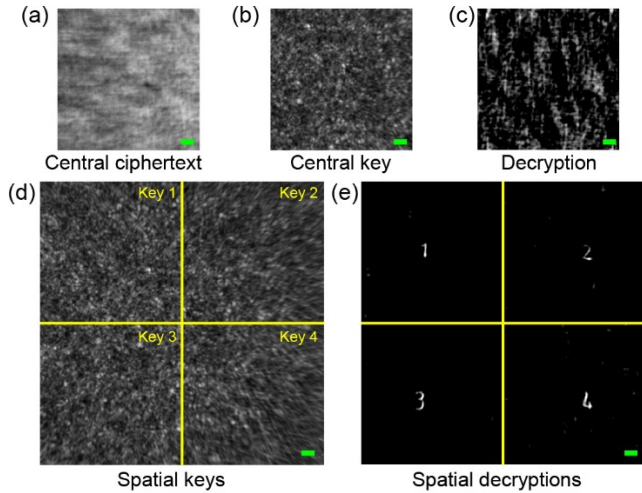

**Fig. 4.** Stitched 2D decryption with various spatial keys. (a-c) Ciphertext, key and decryption with central 220*220 pixels. (d) Various spatial keys with the same size in (a). (e) A full 2D recovery is stitched together with different spatial decryptions. Scale bars: 20 pixels.

In conclusions, we demonstrate a position-multiplexing based cryptosystem to enhance the information security. This method utilizes the two fundamental properties of a scattering medium: memory effect and PSF's spatial decorrelation. Ultra-broadband incoherent illumination and small size of ciphertext image with position multiplexing significantly enhance the security. The unique spatially distributed keys utilized for the decryption are from the same PSF. Multiple plaintexts can be multiplexed in a single ciphertext and sent to all users who will individually decrypted the specific texts with their corresponding keys. As the spatial information of interest are scrambled together or hidden inside the ciphertext, one can decrypt the content with multiple spatial keys but still need to know the keys' order to arrange the multiple pieces of information.

**Funding.** NTU start-up grant (M4081482); Singapore MOE-AcRF Tier-1 (RG70/15); Singapore Ministry of Health National Medical Research Council (NMRC/BNIG/2039/2015).

**Acknowledgement.** We would like to thank the financial supports from NTU start-up grant, Singapore MOE-AcRF Tier-1 grant (RG70/15) and the Singapore Ministry of Health's National Medical Research Council under its <CBRG-NIG (NMRC/BNIG/2039/2015)>